\documentclass[twocolumn,apl,aps,superbib,tightenlines,floatfix,superscriptaddress,showpacs]{revtex4}

\usepackage{amsmath}
\usepackage{graphicx}
\usepackage{amssymb}
\usepackage{color,soul}
\usepackage[T1]{fontenc}
\usepackage{ae,aecompl}
\usepackage{color,soul}

\usepackage{pbox}

\begin{document}

\title{Metastable Memristive Lines for Signal Transmission and Information Processing Applications}%

\author{V. A. Slipko}
\affiliation{Department of Physics and Technology, V. N. Karazin Kharkov National University, Kharkov 61022, Ukraine}
\affiliation{Institute of Physics, Opole University, Opole 45-052, Poland}
\author{Y. V. Pershin}
\email{pershin@physics.sc.edu} \affiliation{Department of Physics and Astronomy and
Smart State Center for Experimental Nanoscale Physics, University of South Carolina, Columbia, South Carolina 29208, USA}
 \affiliation{Nikolaev Institute of Inorganic Chemistry SB RAS, Novosibirsk 630090, Russia}

\begin{abstract}
Traditional studies of memristive devices have mainly focused on their applications in non-volatile information storage and information processing. Here, we demonstrate that the third fundamental component of information technologies -- the transfer of information -- can also be employed with memristive devices. For this purpose,
we introduce a metastable memristive circuit. Combining metastable memristive circuits into a line, one obtains an architecture capable of transferring a signal edge from one space location to another. We emphasize that the suggested transmission lines employ only resistive components. Moreover, their networks (for example, Y-connected lines) have an information processing capability.
\end{abstract}

\maketitle

Currently, the term memristive device~\cite{chua76a} (memristor) is primarily used to denote resistive switching memories that
have been considered the most promising candidates for replacing the state-of-the-art memory technology.
Moreover, it has been established that memristive networks (the networks of memristive devices) are useful to implement
neuromorphic~\cite{pershin09c,prezioso15a}, digital~\cite{borghetti10a,pershin16b} and some unconventional computing architectures~\cite{pershin11d}.
The main advantages of computing with memristors (as well as with memcapacitors~\cite{diventra09a,traversa14a,pershin15a}) are related, in particular, to their ability to store and process information on the same physical platform, massively-parallel dynamics of memristors in networks~\cite{pershin11d}, sub-nanosecond computing times~\cite{Torrezan11a} and low power consumption. Computing with memory circuit elements~\cite{diventra09a} ({\it memcomputing}~\cite{diventra13a}) is thus a promising alternative to the conventional von Neumann computing~\cite{Backus78a}.

The information transfer is another important aspect of modern information technologies. Typically, the signal transmission is considered in the framework of transmission line models having a wide applicability range~\cite{Goleniewski06a,book:1381215,book:450803,Wallraff04a}. The conventional transmission line models employ reactive components -- capacitors and inductors -- for signal transmision. The transmission line losses are taken into account by resistors. Recently, reconfigurable transmission lines utilizing memcapacitors~\cite{diventra09a} instead of capacitors were suggested~\cite{pershin16a}. Transmission characteristics of such lines and thus their functionality can be pre-programmed on demand~\cite{pershin16a}.

\begin{figure}[b]
 \centering{\includegraphics[angle=0,width=0.75\columnwidth]{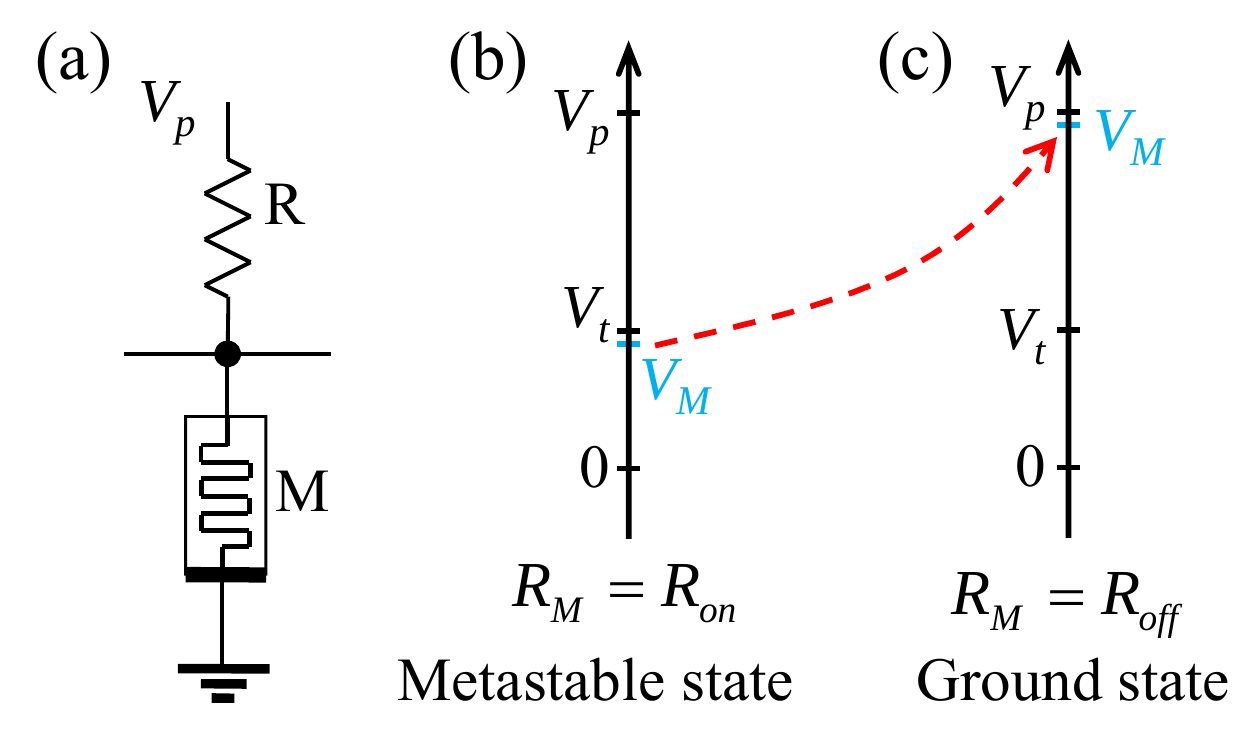}}
\caption{ (a) Metastable memristive circuit. Here, $V_p$ is a power supply voltage. (b) Metastable state of the circuit in (a) is realized when the voltage across M is slightly below its threshold voltage $V_t$. (c) The stable (ground) state corresponds to $R_M=R_{off}$. The red (dashed) line represents the transition from the metastable to the ground state. \label{fig1}}
\end{figure}

The present Letter introduces a different approach to signal transmission uniquely based on the resistive devices. Fig. \ref{fig1}(a) presents its basic block -- a metastable memristive circuit -- combining a resistor R and memristor M. This circuit employs the most common type of memristors characterized by the bipolar threshold-type switching~\cite{pershin11a}. According to the selected connection polarity of M in Fig. \ref{fig1}(a), $R_M$
increases at positive voltages across M, $V_M>V_t$. Here, $R_M$ is the memristance (memory resistance) of M changing between $R_{on}$ and $R_{off}$ (the low- and high-resistance states of memristor), and $V_t$ is the threshold voltage. Moreover, $R$ (the resistance of R in Fig. \ref{fig1}(a) circuit) is selected so that at $R_M=R_{on}$, $V_M$ is slightly below  $V_t$ (see Fig. \ref{fig1}(b)). This circuit configuration can be referred to as a {\it  metastable state}. The circuit can spend an extended time in this state prior being driven out by an input signal or its fluctuation triggering an abrupt (accelerated~\cite{pershin13b}) switching of M. The final state of the circuit (see Fig. \ref{fig1}(c)) is perfectly stable and thus can be referred to as the {\it ground state}.

\begin{figure}[tbp]
 \includegraphics[angle=0,width=\columnwidth]{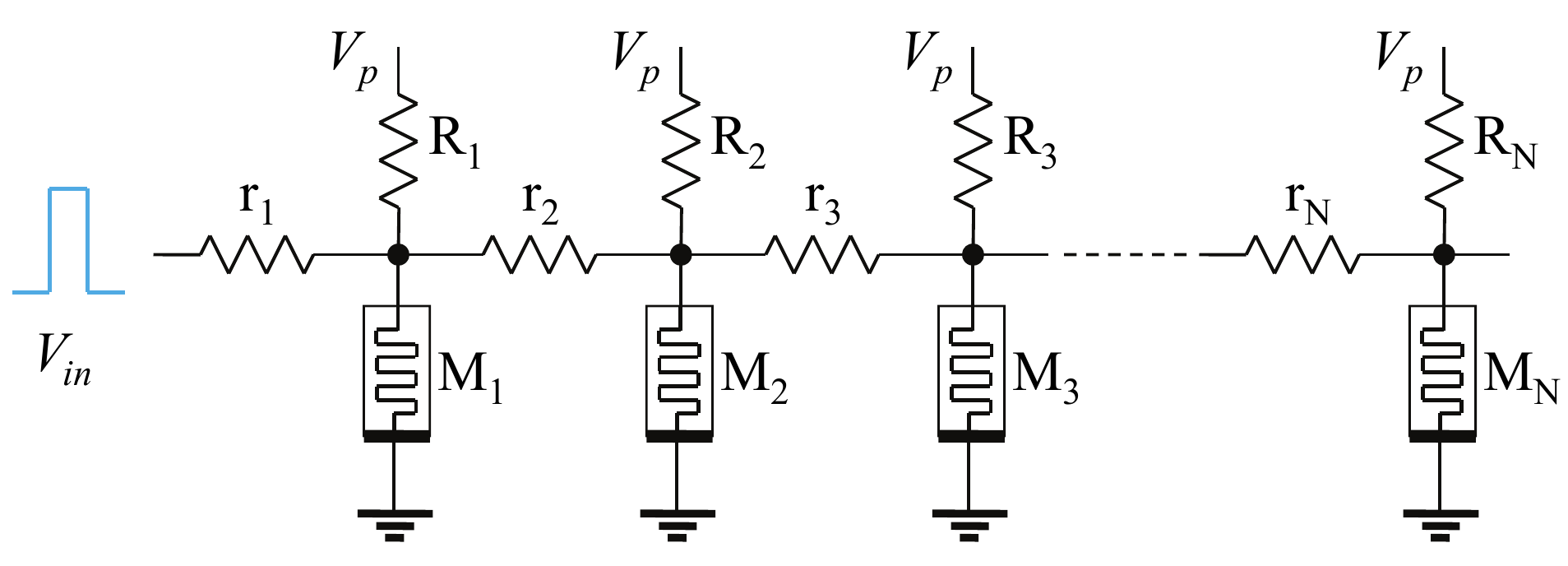}
\caption{ Metastable memristive line consisting of $N$ metastable R-M circuits connected via coupling resistors r$_i$, $i=1,..,N$.
The input signal $V_{in}$" is applied to $r_1$. Note that all the memristors are initialized into the $R_{on}$ state.
 \label{fig2}}
\end{figure}

Next, let us consider a chain (line) of metastable memristive circuits presented in Fig. \ref{fig2}. We argue that under appropriate conditions, this line can transfer a signal edge from one space location to another. Indeed, as it is shown below, there is a certain parameter space where the switching of M$_1$ (from $R_{on}$ to $R_{off}$) initiates the switching of M$_2$, the switching of M$_2$ initiates the switching of M$_3$, and so forth. The applied signal (see Fig. \ref{fig2}), thus, can set off a chain of switching events propagating along the line. We note that since our approach relies on metastable states of memristive devices, such states should be prepared in advance and periodically refreshed (similarly to the laser pumping~\cite{principles} in the area of lasers). Below, we investigate numerically the dynamics of pulse edge propagation along the line and develop a theory of this phenomenon. In particular, we formulate a time-nonlocal equation describing an infinite line and find its solution in a certain limit.

In our calculations, we use the following model of first-order \cite{chua76a} bipolar memristive system with threshold \cite{pershin09b}
\begin{eqnarray}
I=\frac{V}{R_M},~~\frac{\textnormal{d}R_M}{\textnormal{d}t}=f(V,R_M),\;\;\;\;\;\;\;\;\;\;\;\;\;\;\;\;\;
\label{eq:Memristor} \\
f(V,R_M)=\left\{ \begin{array}{cc} \textnormal{sign}(V)
\beta\left( |V|-V_t\right) & \;
\textnormal{if} \; |V|\geq V_t \\ 0 & \textnormal{otherwise}
\end{array} ,
\right. \label{eq:f(V,M)} \\ \nonumber
\end{eqnarray}
where $I$ is the current, $V$ is the voltage, $R_M$ is the memristance changing continuously between $R_{on}$ and $R_{off}$, $\beta$ is the constant defining the rate of change of $R_M$, and $sign\{ ...\}$ is the sign of the argument. According to the above equations, the memristance $R_M$ changes only when $|V|>V_t$, and the direction of change is defined by the device connection and applied voltage polarities.

To describe the line dynamics, we set up a system of equations
based on Kirchoff's rules supplemented by Eqs. (\ref{eq:Memristor})
for the evolution of memristive components.
The voltages across memristive systems M$_i$ are chosen as unknown variables.
The equation for $i$-th metastable circuit ($i=1,2,...,N$) reads
\begin{eqnarray}
\left(\frac{1}{r_{i}}+\frac{1}{r_{i+1}}+\frac{1}{R_{i}}+\frac{1}{R_{M,i}}\right)V_i
-\frac{V_{i-1}}{r_i}-\frac{V_{i+1}}{r_{i+1}}=\frac{V_p}{R_{i}}.
\label{eq:Vi}
\end{eqnarray}
The boundary conditions for the $1$-st and $N$-th  circuits are selected as follows: $V_0=V_{in}$
is the input voltage voltage (see Fig. \ref{fig2}),  and $r_{N+1}=\infty$. Generally,
$2N$ equations  (\ref{eq:Memristor}), (\ref{eq:Vi}) for $2N$ variables
$V_1, V_2,..., V_N$, $ R_{M,1}, R_{M,2},...,  R_{M,N}$ supplemented with initial conditions
(specifically, the initial memristances) fully define the transmission line dynamics.

\begin{figure}[tp]
 \includegraphics[angle=0,width=0.80\columnwidth]{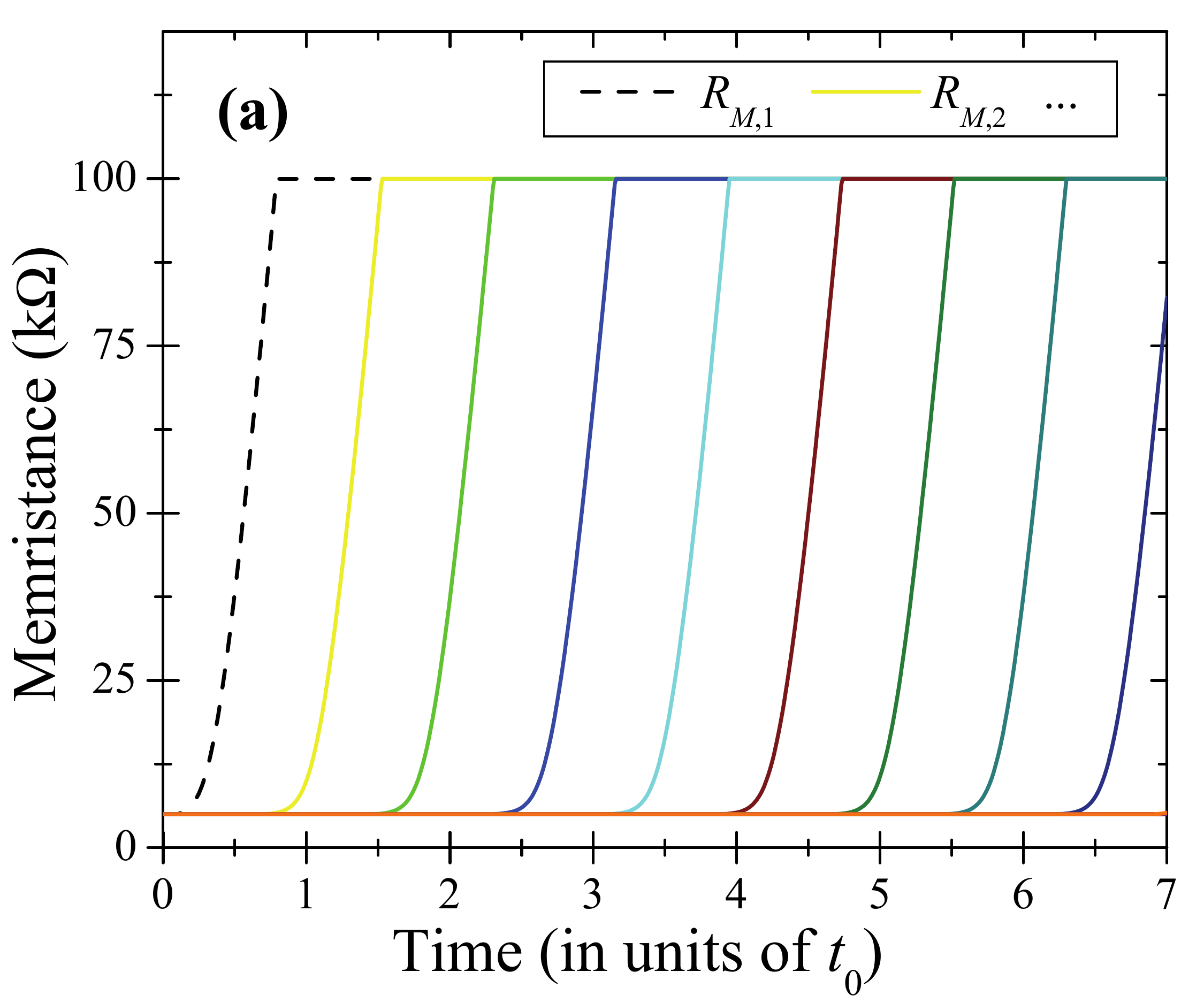}
 \includegraphics[angle=0,width=0.80\columnwidth]{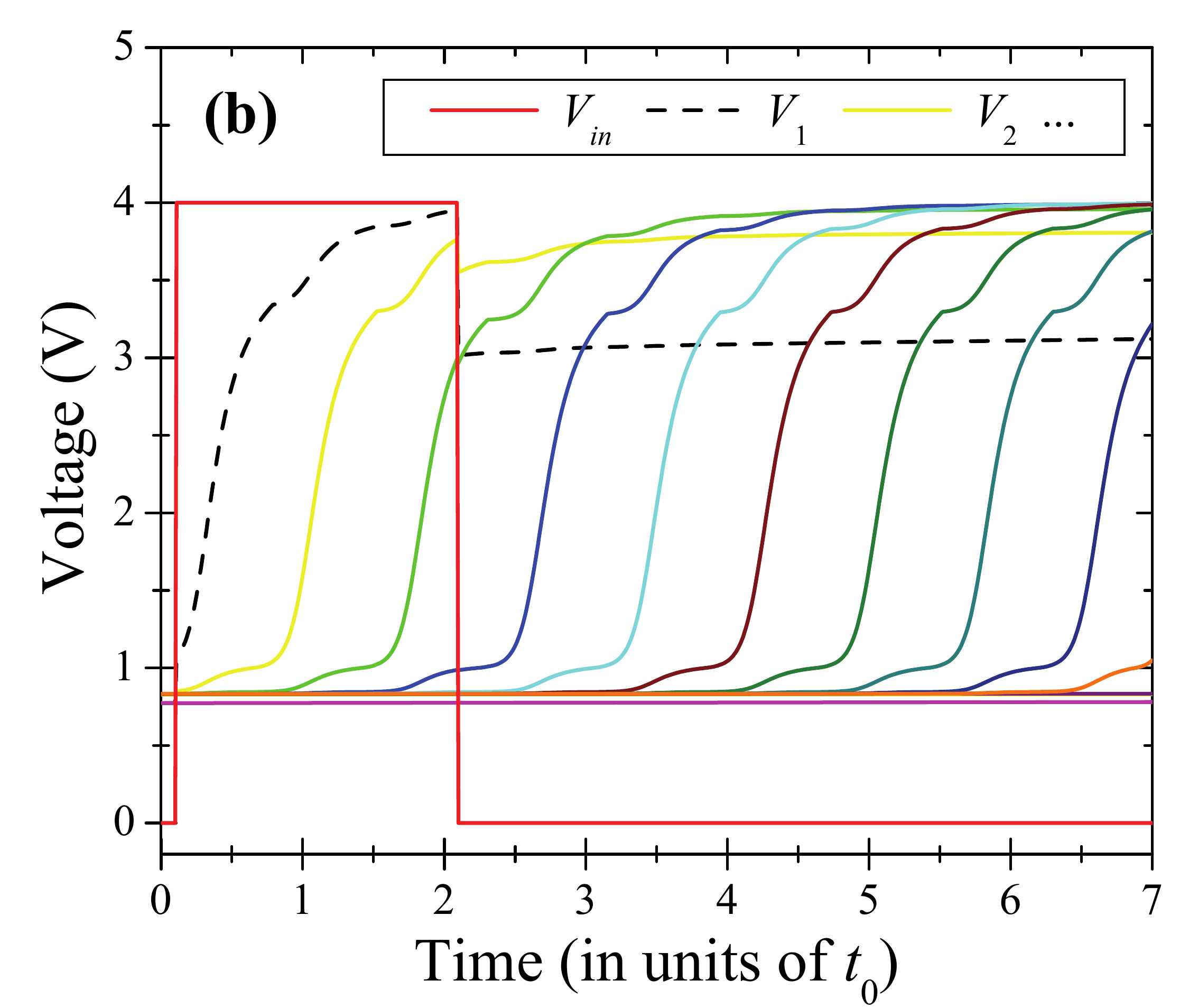}
\caption{ Dynamics of pulse edge propagation in Fig. \ref{fig2} memristive line found as a numerical solution of Eqs. (\ref{eq:Memristor}), (\ref{eq:Vi}). The curves were obtained using the following parameter values: $N=10$, $V_p=5$ V, $r=50$ k$\Omega$, $R=25$ k$\Omega$, $R_{on}=5$ k$\Omega$, $R_{off}=100$ k$\Omega$, $\beta t_0=10^2$ k$\Omega$/(V), $V_t=1$ V and $R_{M,i}(t=0)=R_{on}$. \label{fig3}}
\end{figure}

In what follows we consider a homogeneous metastable memristive line with $r_i=r$,
$R_i=R$, and $R_{M,i}(t=0)=R_{on}$ for $i=1,...,N$. Let us take a closer look at the memristive line dynamics
triggered by a rectangular voltage pulse shown in Fig. \ref{fig3}(b).
Fig. \ref{fig3} presents a numerical solution of the line equations found with a set of parameters specified in
the figure caption. In particular, Fig. \ref{fig3}(a) demonstrates that the switchings of memristors occur
sequentially with almost the same time interval between adjacent switchings. The time dependencies of
voltages (see Fig. \ref{fig3}(b)) are similarly shifted with respect to each other. Their waveforms (neglecting the
boundary effects noticeable in $V_1$ and $V_2$ lines) are essentially the same. Moreover, taking a closer look at any of these voltages, say $V_i$,
one can notice that a slow increase of $V_i$ changes to a fast increase followed by a slow increase. These stages of voltage growth are mainly associated with the switchings of $i-1$, $i$ and $i+1$ memristors, respectively.

It is amply clear that the dynamics in the central part of the line is determined solely by the line properties but not by the boundary conditions (for example, the input pulse waveform or coupling to the external circuit).
Next, we consider the dynamics of pulse edge propagation in the limit of an infinite line as we are not interested in the boundary effects. It is evident that we can safely assume that $V_i$ and $V_{i+1}$ are simply time-shifted with respect to each other, namely, $V_{i+1}(t)=V_{i}(t-\tau)$, where $\tau$ is the pulse edge propagation time per metastable circuit (time interval between adjacent switchings). Then, the system of coupled equations (\ref{eq:Vi}) reduces to a single time-nonlocal equation of the form
\begin{equation}
\left(\frac{2}{r}+\frac{1}{R}+\frac{1}{R_{M,i}}\right)V_i(t)
-\frac{V_i(t-\tau)}{r}-\frac{V_i(t+\tau)}{r}=\frac{V_p}{R}.
\label{eq:V(t)}
\end{equation}
Eq. (4) together with Eq. (\ref{eq:Memristor}) thus describe the pulse edge propagation in an infinite line.  Unfortunately, even the set of Eqs.  (\ref{eq:Memristor}), (\ref{eq:V(t)}) is rather complicated one, not only
because Eq. (\ref{eq:V(t)}) includes retarded and advanced times, but also
because it is nonlinear. Even  for one of the simplest possible $f(V,M)$ given by Eq. (\ref{eq:f(V,M)}) we cannot solve the system
 (\ref{eq:Memristor}), (\ref{eq:V(t)}) analytically.

In fact, an approximate solution to the problem can be obtained in the limit of independent dynamics. In this limit, we assume that at each instant of time only one memristor is changing its state (switching). To proceed, let's focus on Eqs. (\ref{eq:Vi}) for $(i-1)$, $i$-th and $(i+1)$ metastable circuits:
\begin{eqnarray}
\left(\frac{2}{r}+\frac{1}{R}+\frac{1}{R_{M,i-1}}\right)V_{i-1}-\frac{V_{i-2}}{r}-\frac{V_{i}}{r} &=& \frac{V_p}{R},
\label{eq:Vi-1}\\
\left(\frac{2}{r}+\frac{1}{R}+\frac{1}{R_{M,i}}\right)V_{i}-\frac{V_{i-1}}{r}-\frac{V_{i+1}}{r} &=& \frac{V_p}{R},
\label{eq:Vi2}
\\
\left(\frac{2}{r}+\frac{1}{R}+\frac{1}{R_{M,i+1}}\right)V_{i+1}-\frac{V_{i}}{r}-\frac{V_{i+2}}{r} &=& \frac{V_p}{R}.
\label{eq:Vi+1}
\end{eqnarray}
Next, the following approximations are made: (i) the voltages at $(i-2)$ and $(i+2)$ nodes are replaced by some constant values, $V_{i-2}=V_{off}$, $V_{i+2}=V_{on}$, and (ii) it is assumed that $(i-1)$ and $(i+1)$ memristors are in the $R_{off}$ and $R_{on}$ states, respectively, that is $R_{M,i-1}=R_{off}$ and $R_{M,i+1}=R_{on}$. Here, $V_{on}$ and $V_{off}$ are given by
\begin{eqnarray}
V_{on(off)}=\frac{R_{on(off)}}{R_{on(off)}+R}V_p
\label{eq:Vonf}
\end{eqnarray}
representing the voltages in the line with all memristors in either $R_{on}$ or $R_{off}$ state. The above approximations make possible to truncate the system of equations (\ref{eq:Vi}).

From the truncated system of equations (\ref{eq:Vi-1})-(\ref{eq:Vi+1}) we find
\begin{eqnarray}
V_i(R_{M,i})=V_p \frac{Y_1 R_{M,i}}{Y_2R_{M,i}+1},
\label{eq:Vi(M)}
\end{eqnarray}
\begin{eqnarray}
V_{i+1}(R_{M,i})= \frac{V_i(R_{M,i})}{rY_{on}}+\gamma_{on}V_p,
\label{eq:Vi+1(M)}
\end{eqnarray}
\begin{eqnarray}
V_{i-1}(R_{M,i})=\frac{V_i(R_{M,i})}{rY_{off}}+\gamma_{off}V_p,
\label{eq:Vi-1(M)}
\end{eqnarray}
where
\begin{eqnarray}
Y_1=\frac{1}{R}+\frac{r(R+R_{off})+RR_{off}}{r^2R(R+R_{off})Y_{off}}\nonumber \;\;\;\;\;\;\;\;\;\;\;\;\;\;\;\;\;\;          \\
+\frac{r(R+R_{on})+RR_{on}}{r^2R(R+R_{on})Y_{on}},
\label{eq:Y1}
\end{eqnarray}
\begin{eqnarray}
Y_2=\frac{2}{r}+\frac{1}{R}-
\frac{1}{r^2}\left(\frac{1}{Y_{on}}+\frac{1}{Y_{off}}\right),
\label{eq:Y2}
\end{eqnarray}
\begin{eqnarray}
Y_{on(off)}=\frac{2}{r}+\frac{1}{R}+\frac{1}{R_{on(off)}},
\label{eq:Yon}
\end{eqnarray}
and
\begin{eqnarray}
\gamma_{on(off)}=\frac{r(R+R_{on(off)})+RR_{on(off)}}{rR(R+R_{on(off)})Y_{on(off)}}.
\label{gamma}
\end{eqnarray}

\begin{figure}[t]
 \includegraphics[angle=0,width=0.85\columnwidth]{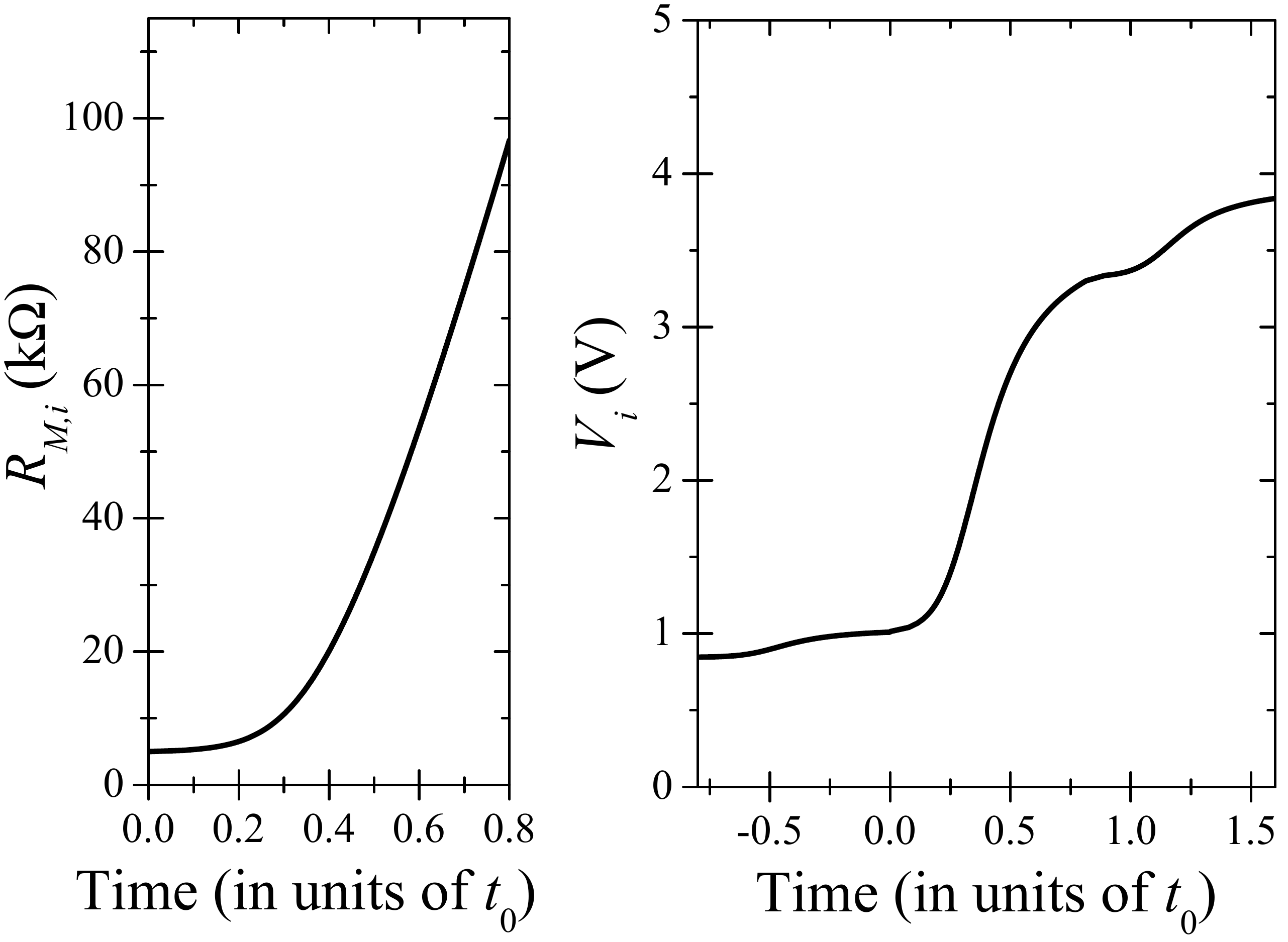}
\caption{ Time dependencies of the memristance, $R_{M,i}$, and voltage across the memristor, $V_i$, found using the analytical model described in the text. This plot was obtained using the same set of parameters as those detailed in Fig. \ref{fig3}. \label{fig4}}
\end{figure}

Now taking into account Eq. (\ref{eq:Vi(M)}) and using the initial condition $M_i(t=0)=R_{on}$ we integrate Eq. (\ref{eq:Memristor})
and obtain implicitly the time dependence of $R_{M,i}(t)$:
\begin{eqnarray}
t=\frac{1}{\beta}\left\{\frac{Y_2(R_{M,i}-R_{on})}{Y_1 V_p-Y_2V_t}\right.\nonumber  \qquad \qquad \qquad  \qquad \qquad \qquad  \\
\left.+\frac{Y_1V_p}{(Y_1 V_p-Y_2V_t)^2}
\ln \frac{(Y_1 V_p-Y_2V_t)R_{M,i}-V_t}{(Y_1 V_p-Y_2V_t)R_{on}-V_t}\right\}. \;\;
\label{eq:Mi(t)}
\end{eqnarray}
We note that Eq. (\ref{eq:Mi(t)}) can be used to find the switching time $T$ of every memristor in the line. For this purpose, one should just substitute $R_{M,i}=R_{off}$ in
the right-hand side of Eq. (\ref{eq:Mi(t)}). We have found a very good agreement between the results of exact numerical calculations and analytical solution (\ref{eq:Vi(M)}),(\ref{eq:Mi(t)}). Fig. \ref{fig4} presents $R_{M,i}(t)$ and $V_i(t)$ found using the analytical model.
Eqs. (\ref{eq:Vi(M)})  and (\ref{eq:Mi(t)}) determine the time dependence of voltage across the switching memristor.
The pairs of Eqs. (\ref{eq:Vi+1(M)}) , (\ref{eq:Mi(t)}) and Eqs. (\ref{eq:Vi-1(M)}) , (\ref{eq:Mi(t)}) can be used to obtain the
voltage across $i$-th memristor in the situation when $(i-1)$ memristor is switching, and, correspondingly, when $(i+1)$ memristor is switching.

In the above consideration, the $i$-th memristor starts switching at $t=0$. Therefore,
the time $\tau$ it takes for the switching edge to move from $i$th to $(i+1)$ metastable circuit along the line
can be found as the time when a suitable condition to start the switching of $(i+1)$
memristor is established, namely, when
\begin{eqnarray}
V_{i+1}(R_{M,i}^{\tau}))=V_t.
\label{eq:tau}
\end{eqnarray}
By using Eqs. (\ref{eq:tau}), (\ref{eq:Vi+1(M)}), and (\ref{eq:Vi(M)}) we
determine the value of $R_{M,i}^\tau$  such that the condition (\ref{eq:tau}) is satisfied:
\begin{eqnarray}
R_{M,i}^\tau=\frac{rY_{on}(V_t-\gamma_{on}V_p)}{Y_2rY_{on}(V_t-\gamma_{on}V_p)-Y_1
V_p}.
\label{eq:Mitau}
\end{eqnarray}
Plugging this value of $R_{M,i}$ into the right-hand side of Eq. (\ref{eq:Mi(t)}) we get $\tau$.
Note that the switching time $T$ is always longer than the pulse edge propagation time per metastable circuit, $\tau$.

\begin{figure}[tbp]
 \includegraphics[angle=0,width=0.85\columnwidth]{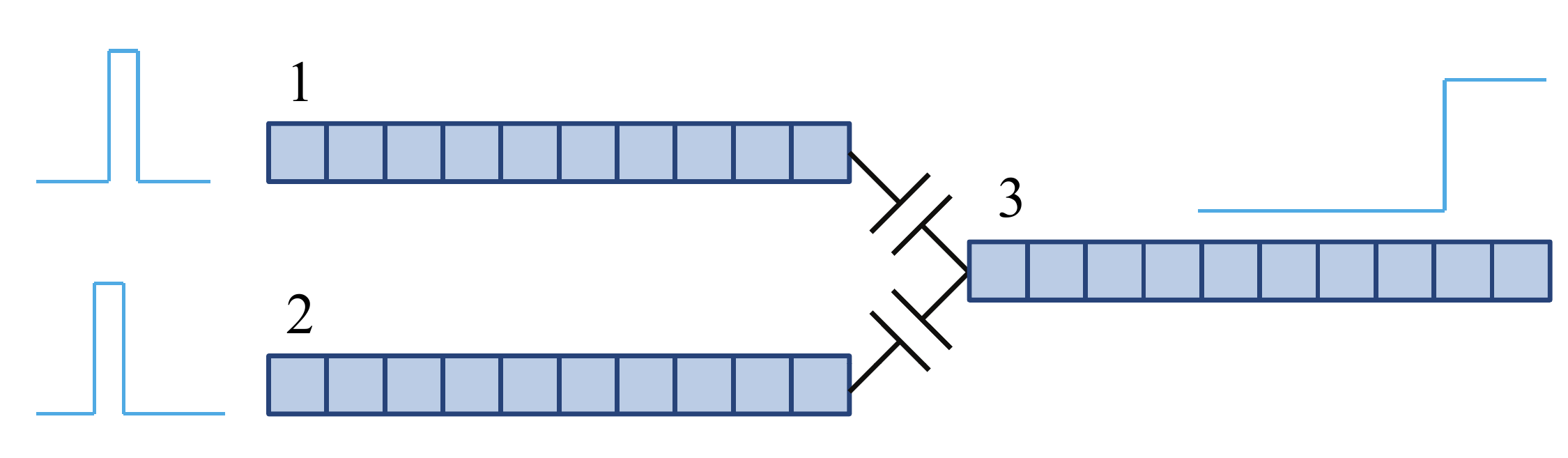}
\caption{{Boolean logic with metastable memristive lines. This plot conceptually presents two input (1,2) and one output (3) lines coupled capacitively.
Depending on the coupling strength, either one or two simultaneously propagating pulse edges in the input lines are required to induce the switching in the output line. One can interpret such a functionality as an OR or AND gate, respectively. Note that a resistive coupling between the lines can also result in a similar functionality.} \label{fig5}}
\end{figure}

In conclusion, in this Letter we introduced metastable memristive states, circuits and lines. The signal transmission through metastable transmission lines was investigated using both numerical and analytical approaches. An approximate analytical solution was found in the framework of a single memristor switching approximation. Thus, we have established an innovative approach to signal transmission, which is unique in being based on only resistive components. Moreover, one can also envisage purely capacitive transmission lines, where the capacitive components replace the corresponding resistive ones. However, this idea needs further investigation.

Furthermore, we note that metastable memristive lines can also find applications in the area of information processing. For example, the time delays introduced by these lines could be of use in the development of race logic architectures~\cite{madhavan2015race}. Moreover, capacitively Y-connected lines (see Fig. \ref{fig5} for an example) are capable to implement some boolean logic operations, such as AND and OR.
Some additional information regarding this idea is provided in Fig. \ref{fig5} caption.

This work has been partially supported by the Russian Scientific Foundation
grant No. 15-13-20021. VAS acknowledges the support by the Erasmus Mundus Action 2 ACTIVE programme (Agreement No. 2013-2523/001-001 EMA2).

\bibliography{memcapacitor}

\end{document}